\documentclass[runningheads]{llncs}
\usepackage{graphicx}
\usepackage[utf8]{inputenc}
\usepackage{lipsum}
\usepackage{tabularx}
\usepackage[portuguese]{babel}
\usepackage{hyperref}
\usepackage[bottom]{footmisc}
\usepackage{cite}
\usepackage{microtype}

\begin{document}
\title{Análise de Segurança Baseada em Roles para Fábricas de Software}
\author{Miguel Loureiro\inst{1}
\and
Luísa Lourenço\inst{2}
\and
Lúcio Ferrão\inst{2}
\and
Carla Ferreira\inst{1}
}
\authorrunning{M. Loureiro et al.}

\institute{NOVA-LINCS \& DI, FCT, Universidade Nova de Lisboa \and
OutSystems Portugal}
\maketitle              
\begin{abstract}

A maioria das fábricas de software contêm aplicações com informação sensível que necessita de ser protegida contra quebras de confidencialidade e integridade, as quais podem ter consequências graves. No contexto de fábricas de grandes dimensões com aplicações complexas, não é viável analisar  manualmente acessos a informação sensível sem os mecanismos de segurança necessários.

Este artigo apresenta uma técnica de análise estática de segurança para fábricas de software, baseada em políticas de segurança com \textit{roles}. 
Na abordagem proposta é construída uma representação em forma de grafo da fábrica de software, tendo por base a política de segurança definida pelo utilizador. Posteriormente o modelo é analisado para encontrar acessos a informação onde a política de segurança é quebrada, garantindo que todos os possíveis estados de execução são analisados.

Foi desenvolvida uma \textit{proof of concept} que implementa a nossa técnica de análise para fábricas de software OutSystems. Com base nos resultados da análise são gerados relatórios de segurança, de modo a dar visibilidade às falhas encontradas e ajudar a equipa de desenvolvimento a prioritizar estas falhas. O protótipo foi avaliado usando fábricas de software de grandes dimensões e com requisitos fortes de segurança. Foram encontradas várias falhas de segurança, algumas graves, que dificilmente seriam encontradas sem a nossa análise.

\end{abstract}
\section{Introdução}

Na informática, a segurança é uma área em constante evolução, de importância cada vez maior à medida que a nossa sociedade se vai tornando cada vez mais avançada e dependente em infraestructuras como a internet. Relatórios de segurança recentes\footnote{Relatórios de segurança da Symantec 2017 e 2018: \url{https://www.symantec.com/content/dam/symantec/docs/reports/istr-22-2017-en.pdf} e \url{https://www.symantec.com/security-center/threat-report}} mostram que em 2017 houve um aumento de 13\% no número de vulnerabilidades de segurança encontradas. Outro relatório de segurança\footnote{Relatório de seguraça Trustwave 2017: \url{https://www2.trustwave.com/2017-Trustwave-Global-Security-Report.html/}} mostra que mais de metade das vulnerabilidades de segurança encontradas em aplicações web fazem parte da categoria de fugas de informação.

Vulnerabilidades de segurança, nomeadamente fugas de informação, têm  repercussões negativas tanto para clientes como para empresas.
Expor a informação pessoal de um indivíduo pode ter efeitos graves. Por exemplo, se a informação bancária de um cliente for exposta publicamente, é um alvo para roubo, ou se informação pessoal sensível for exposta pode ser vítima de roubo de identidade.
Ao nível empresarial, fugas de informação reduzem o nível de confiança dos clientes, afetam o seu valor de mercado, e criam complicações legais.

Este trabalho foi desenvolvido com o objetivo de melhorar a segurança de fábricas de software, \emph{i.e.}, um conjunto de aplicações que podem estar relacionadas e nas quais pode existir troca de informação.
Em fábricas de software compostas por um número elevado de aplicações web que trabalhem com informação sensível, torna-se difícil assegurar que não existem falhas de segurança. 
Devido à quantidade de aplicações e à complexidade de cada uma delas, é complicado determinar que informação é exposta para cada um dos \textit{roles} em cada \textit{entrypoint} (\emph{e.g.}, ecrãs, \textit{endpoints} REST). 
Assim, torna-se difícil colocar os devidos mecanismos de segurança para proteger informação sensível, e ainda mais difícil determinar à posteriori onde faltam estes mecanismos de segurança.

Surge assim um problema para equipas que desenvolvam fábricas de software de grandes dimensões. É inevitável que, por vezes, os programadores se esqueçam de proteger informação sensível com mecanismos de segurança. 
Assim, existe uma propagação de falhas de segurança, que dificilmente será manualmente detetadas posteriormente.
Portanto, é inexequível garantir a segurança de uma fábrica de software sem uma análise de segurança que detete falhas. 

As contribuições chave apresentadas neste artigo são uma técnica de análise estática de segurança para fábricas de software compostas por várias aplicações web com um modelo de controlo de acesso baseado em \textit{roles}.
A análise recebe uma política de segurança que associa recursos a \textit{roles} e encontra, analisando todas as aplicações da fábrica, \textit{entrypoints} e sub-rotinas chamadas, onde a política é quebrada.

A técnica foi implementada numa ferramenta \textit{proof of concept} direcionada para fábricas de software OutSystems. Depois de detetar falhas de segurança numa fábrica, a ferramenta gera  um relatório de segurança que, ao dar visibilidade a estas falhas, ajuda o utilizador a encontrar e prioritizar falhas na sua fábrica de software.

O resto do documento está estruturado da seguinte forma. Na Secção~\ref{sec:solution} é feita a descrição geral da técnica de análise.
Na Secção~\ref{sec:ferramenta} é apresentada a descrição da implementação da técnica numa \textit{proof of concept}.
A descrição da avaliação à técnica desenvolvida é apresentada na  Secção~\ref{sec:eval}. De seguida, a Secção~\ref{sec:related} descreve o trabalho 
relacionado à técnica desenvolvida. Por último, a Secção~\ref{sec:conc} apresenta as conclusões e observações finais.

\section{Visão Geral da Análise}
\label{sec:solution}

Nesta secção é feita uma descrição geral da técnica de análise estática desenvolvida. 
São feitas as seguintes assunções sobre a arquitetura duma aplicação web genérica que utilize um modelo de controlo de acessos baseado em roles: 

\begin{itemize}
    \item Os utilizadores do sistema têm um  ou mais \textit{roles} atribuídos; 
    \item O acesso a cada \textit{entrypoint} pode estar limitado a um ou mais \textit{roles};  
    \item O modelo contém primitivas para atribuir, remover e verificar \textit{roles}.
\end{itemize}

\begin{figure}[!t]
    \centering
    \includegraphics[width=.4\linewidth]{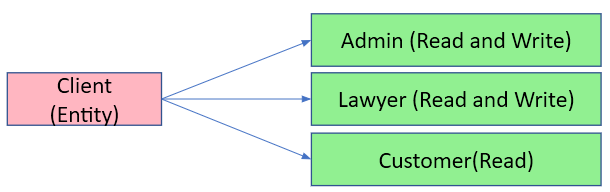}
    \caption{Exemplo de uma política de segurança}
    \label{fig:policy}
\end{figure}

A nossa técnica de análise tem por base numa política de segurança definida pelo utilizador. Cada uma das regras da política de segurança  consiste num recurso (\emph{e.g.}, uma tabela de uma base de dados) e nos conjuntos de \textit{roles} com acesso de leitura e de escrita. A Figura~\ref{fig:policy} mostra um exemplo duma política de segurança, que define que o recurso (entidade) \textit{Client} pode ser lido e escrito pelos \textit{roles} \textit{Admin} e \textit{Lawyer}, e pode ser lido pelo \textit{role} \textit{Customer}.

O processo de análise começa por produzir uma síntese da fábrica de software com base na política de segurança recebida. Tendo em conta a dimensão das  fábricas de software e à quantidade de elementos diferentes e complexos que cada fábrica pode ter, é importante construir um modelo sintetizado, só com informação relevante, sob o qual se pode realizar a análise. 

O modelo sintetizado duma fábrica de software é construído num call graph, um grafo que representa as relações de chamada entre sub-rotinas dum programa. Este call graph contém:
\begin{itemize}
    \item Os \textit{entrypoints} das várias aplicações da fábrica;
    \item As sub-rotinas chamadas por cada \textit{entrypoint} da fábrica;
    \item As primitivas de leitura e escrita de recursos da fábrica;
    \item Os recursos da fábrica.
\end{itemize}
Um call graph apenas com esta informação já daria resultados úteis, mas devido à falta de alguma informação relevante -- como as chamadas de primitivas que atribuem, retiram ou verificam os \textit{roles} de um utilizador 
-- ia resultar num elevado número de falsos positivos. Como tal, os nós do call graph associados a sub-rotinas guardam também o control flow graph (CFG) dessa sub-rotina.

Depois de construir o modelo sintetizado, o modelo é analisado à procura de falhas de segurança. Esta análise é composta por três passos:
\begin{enumerate}
    \item Procurar \textit{potenciais} falhas de segurança no call graph;
    \item Expandir o call graph de cada \textit{potencial} falha, usando os CFGs das sub-rotinas;
    \item Analisar o CFG de cada \textit{potencial} falha, para validar a existência de uma falha.
\end{enumerate}
O primeiro passo consiste em utilizar o call graph para encontrar todas as potenciais falhas de segurança. 
Uma \emph{potencial} falha é um caminho entre um recurso \textit{X} que faça parte da política de segurança e um \textit{entrypoint} \textit{Y} que permita o acesso a um utilizador com um \textit{role} que não faça parte da política de \textit{X}.
É de notar que não é possível determinar de forma definitiva que os caminhos identificados representam falhas reais,  pois o call graph não tem informação completa, como primitivas que verificam se o utilizador atual possuí um \textit{role} (\textit{check role}). A Figura~\ref{fig:ex_checkrole} mostra um exemplo da utilização de uma verificação de \textit{role} para proteger o acesso a um recurso -- só se o utilizador tiver o \textit{role} \textit{Admin} é que a operação de escrita \textit{DeleteLegalCase} vai ser chamada.
Se um recurso identificado como uma potencial falha estiver protegido por um \textit{check role} usado num bloco condicional, então trata-se dum falso positivo.
Para encontrar todas as potenciais falhas, é utilizado um algoritmo de pesquisa em profundidade.

O segundo passo consiste em expandir o call graph de cada potencial falha com o CFG interno de cada sub-rotina. 
Obtém-se assim um CFG, com mais informação em relação ao call graph.

O terceiro passo consiste em analisar o CFG resultante do segundo passo para validar a potencial falha encontrada.
Esta análise é feita com um algoritmo de pesquisa em profundidade eficiente\footnote{Não percorre caminhos redundantes.}. 
A análise dum CFG começa no nó do \textit{entrypoint} e é mantido um estado durante a análise de cada caminho. 
Este estado vai conter os \textit{roles} que o utilizador \textit{pode} ter e os \textit{roles} que o utilizador \textit{tem}. 

O estado inicial é determinado a partir do conjunto de roles com acesso ao \textit{entrypoint}, sendo atualizado durante a travessia com as primitivas que atribuem, retiram ou verificam um \textit{role} ao utilizador. 
Um dos mecanismos de segurança usados consiste na utilização da primitiva \textit{check role} dentro de um bloco condicional. O estado vai ser alterado para cada ramo do bloco condicional, consoante a condição usada.
Ao passar por um nó dum recurso que pertença à política de segurança, verifica-se o estado para determinar se existe uma falha de segurança. 
Se o utilizador não possuir um dos \textit{roles} da política, e puder ter um \textit{role} que não faça parte da política, trata-se duma potencial falha de segurança que deve ser reportada.

O resultado da análise é um conjunto de \textit{entrypoints} com falhas de segurança, as sub-rotinas chamadas por cada \textit{entrypoint} com falhas de segurança, e os caminhos tanto no call graph como no control flow graph de cada falha encontrada. 

Na próxima secção é descrito como é que os resultados da análise são usados para ajudar o utilizador final a visualizar e a prioritizar as falhas encontradas.

\section{Ferramenta de Análise e Relatórios de Segurança}
\label{sec:ferramenta}

Nesta secção é descrita a implementação da técnica de análise de segurança e a geração de relatórios com o resultado da análise. 
A implementação da técnica foi direcionada a fábricas de software OutSystems, e para tal foi feito um mapeamento dos elementos do modelo genérico da análise para elementos do modelo OutSystems. 

A OutSystems\footnote{\url{https://www.outsystems.com/}} é uma plataforma que permite o desenvolvimento rápido de aplicações web e mobile.
As aplicações OutSystems podem ter vários  \textit{entrypoints}, e.g., ecrãs, \textit{endpoints} REST, \textit{endpoints} SOAP.
Um ecrã é um elemento do modelo OutSystems que contém uma interface para utilizador final interagir com a aplicação e que é usado como \textit{entrypoint} da aplicação.

Para simplificar, a nossa análise só contempla ecrãs como \textit{entrypoints} de aplicações. 
Um ecrã em OutSystems tem um conjunto de permissões, ou seja, um conjunto de \textit{roles} com acesso ao ecrã. 
Em OutSystems, os dados (sensíveis ou não) duma aplicação estão armazenados em entidades. 
Pode-se assumir que uma entidade OutSystems é uma tabela numa base de dados SQL.
O modelo OutSystems possuí três primitivas de RBAC -- \textit{GrantRole}, \textit{RevokeRole} e \textit{CheckRole} -- que respetivamente atribuem um \textit{role} ao utilizador, retiram um \textit{role} ao utilizador, e verificam se o utilizador tem um \textit{role}.

\begin{figure}[!t]
    \centering
    \includegraphics[width=0.6\linewidth]{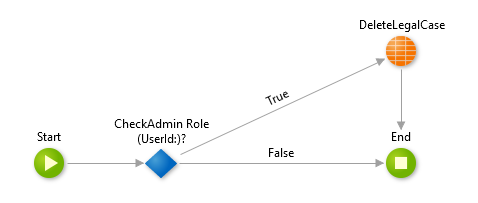}
    \caption{Exemplo de \textit{CheckRole} a proteger o acesso de escrita a uma  entidade \textit{LegalCase}}
    \label{fig:ex_checkrole}
\end{figure}

Em OutSystems é comum utilizar a primitiva \textit{CheckRole} para proteger o acesso a entidades.
A Figura~\ref{fig:ex_checkrole} contém um exemplo dum \textit{CheckRole} dentro dum bloco condicional a garantir que o acesso à entidade \textit{LegalCase} é restrito a utilizadores com o role \textit{Admin}. 

O caminho duma potencial falha de segurança  começa num ecrã e termina numa entidade. 
Um ecrã possuí ações que podem aceder a entidades, ou podem chamar outras ações que eventualmente acedem a uma entidade.
Mais precisamente, uma potencial falha de segurança consiste num caminho entre uma entidade que faz parte da política de segurança, um ecrã que permite a entrada a \textit{roles} que não fazem parte da política de segurança da entidade, e um caminho sem mecanismos de segurança até à entidade.

No desenvolvimento da ferramenta foi implementada uma interface para ajudar os 
utilizadores a definirem as suas políticas de segurança. 
A interface mostra uma lista de todas as entidades e \textit{roles} da fábrica de software e permite o utilizador definir que \textit{roles} têm acesso de leitura e escrita a cada entidade.
Cada regra duma política de segurança associa a cada entidade dois conjuntos de \textit{roles}, os \textit{roles} com acesso de leitura e os \textit{roles} com acessos de escrita.

Após a definição da  política de segurança, pode-se executar a análise, cujo resultado vai ser uma coleção de ecrãs com (potenciais) falhas de segurança, cada um com uma coleção de ações com falhas, mais os respetivos caminhos do call graph e CFG.

\begin{figure}[t]
    \centering
    \includegraphics[width=0.9\linewidth]{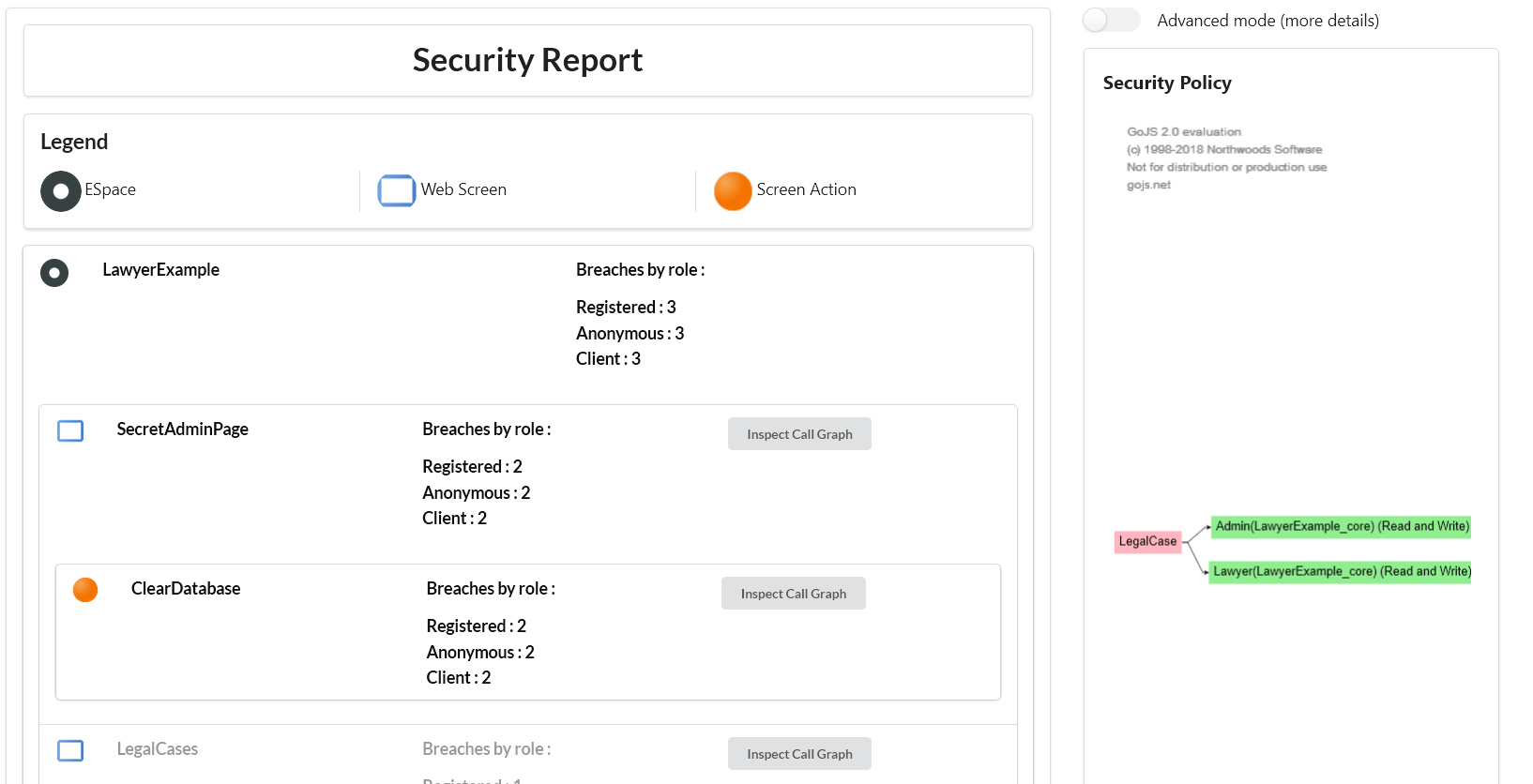}
    \caption{Relatório de segurança}
    \label{fig:report1}
\end{figure}

Para que resultados da análise sejam úteis para o utilizador final, a ferramenta constrói também um relatório de segurança.
O relatório vai mostrar todas as falhas encontradas, ordenando-as por risco de modo a ajudar a sua prioritização. A Figura~\ref{fig:report1} mostra um relatório de segurança produzido pela ferramenta. 
Este relatório começa por listar todos os módulos de aplicação com potenciais falhas de segurança. 
Em cada entrada do relatório é também mostrado o número de falhas encontradas e que \textit{roles} é que quebraram a política de segurança. Ao selecionar uma entrada do relatório, essa é expandida e são mostrados todos os ecrãs desse módulo que contêm falhas. 
Um entrada dum ecrã também pode ser expandida, mostrando todas as ações com falhas encontradas nesse ecrã. A Figura~\ref{fig:report1} mostra que foram encontradas falhas no módulo ``LaywerExample'', mais especificamente no seu ecrã ``SecretAdminPage''.
Neste ecrã, para cada um dos roles \textit{Registered}, \textit{Anonymous} e \textit{Client}, foram encontradas duas potenciais falhas de segurança, onde utilizadores com esses roles podem ter acesso indevido a uma entidade.

\begin{figure}[t]
    \centering
    \includegraphics[width=.9\linewidth]{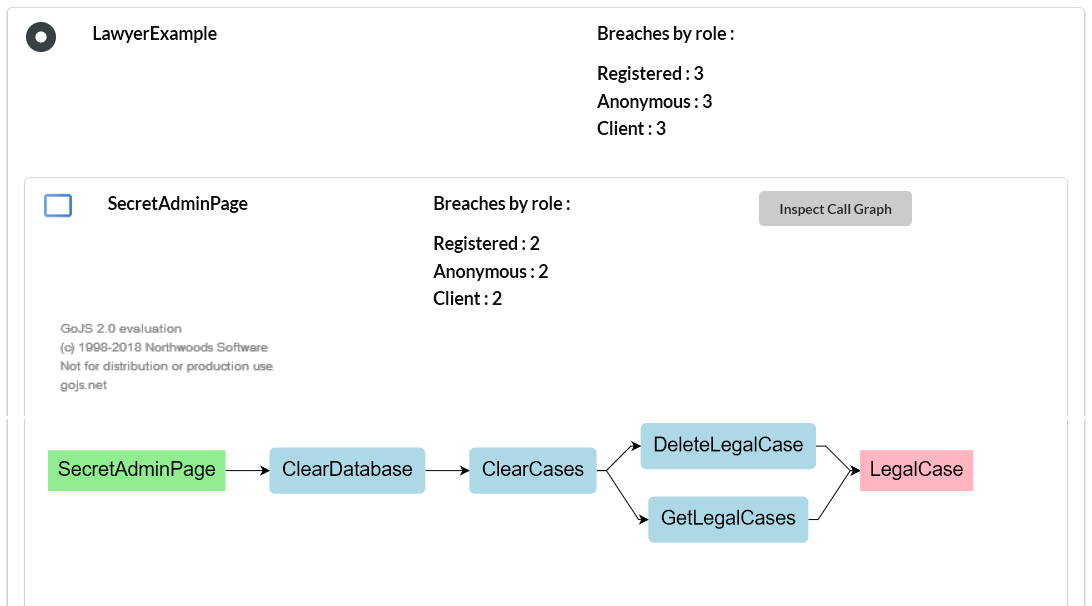}
    \caption{Relatório de segurança, com call graph expandido}
    \label{fig:report2}
\end{figure}

Nas entradas de ecrãs e ações, é também possível selecionar no botão ``Inspect Call Graph'', que vai mostrar o call graph da(s) falha(s) encontradas naquele ecrã/ação. A Figura~\ref{fig:report2} mostra esta funcionalidade do relatório de segurança.

\section{Avaliação}
\label{sec:eval}

Esta secção descreve a avaliação empírica da ferramenta que implementa a análise para fábricas de software OutSystems, descrita na Secção~\ref{sec:ferramenta}. 

Para a avaliação foram utilizadas três fábricas de software. A primeira fábrica é um \textit{mock-up} que consiste num exemplo simples dum sistema de gestão para uma firma de advogados. 
As outras duas fábricas utilizadas, A e B, são fábricas de clientes da OutSystems~\footnote{Não revelados por motivos de privacidade.}. A utilização de uma fábrica \textit{mock-up} é útil porque é possível introduzir falhas de segurança e verificar se a análise é capaz de as detetar. 
No entanto não é suficiente, não só devido à sua reduzida dimensão mas também porque numa fábrica real podem-se encontrar falhas que dificilmente seríamos capazes de reproduzir. 

A fábrica \textit{mock-up} tem dois módulos, um para a UI da aplicação e outro para o modelo de dados. 
O modelo de dados é simples e contém três entidades: \textit{Client}, \textit{Lawyer} e \textit{LegalCase}.
\textit{Client} representa os clientes da firma de advogados, \textit{Lawyer} representa os advogados da firma, e \textit{LegalCase} representa os processos legais da firma.
Cada processo está associado a um cliente e a um advogado.

Nesta fábrica foram introduzidas algumas falhas de segurança, por exemplo, ecrãs (\textit{entrypoints}) que dão acesso a \textit{roles} que não deviam e que chamam ações (sub-rotinas) que acedem a entidades sensíveis sem \textit{checks} de segurança.

Em relação às fábricas de software dos clientes, a fábrica A é de tamanho médio, com cerca de 30 módulos, tendo mais de 200 ecrãs (\textit{entrypoints}), 200 entidades e 35 \textit{roles}. 
Esta fábrica contém várias aplicações usadas internamente por uma empresa.

A fábrica B é uma das maiores fábricas existentes em OutSystems. 
Contém mais de 700 módulos de aplicação, com mais de 2000 entidades, 2500 ecrãs e 321 \textit{roles}. 
A fábrica B contém um sistema semelhante ao \textit{mock-up} da firma de advogados, mas direcionado para uma área profissional diferente. 
Esta fábrica vai ser o nosso \textit{benchmark} para performance e escalabilidade. 
Ou seja, se a nossa técnica tiver boa performance com esta fábrica podemos assumir que a sua performance também será boa para a maioria das fábricas de software existentes. 

\subsection{Validação}

Para validar a nossa técnica, testámos-la com as três fábricas de software. 
Utilizando a nossa política de segurança para a fábrica \textit{mock-up}, os resultados obtidos foram todos corretos. A análise foi capaz de detetar todas as falhas de segurança introduzidas, sem falsos positivos nem falsos negativos.

Em relação às fábricas de software A e B, seria virtualmente impossível validá-las completamente, devido às suas dimensões, falta de contexto, complexidade de políticas de segurança e quantidade de resultados em cada relatório. 
Foi feita uma amostragem de resultados de análises às fábricas A e B, que foi validada manualmente.

Para validar a nossa análise das fábricas A e B, obtivemos algumas políticas de segurança dos clientes, e também definimos algumas políticas nossas. Para as regras que definimos, escolhemos entidades que incluíam informação sensível e declaramos que os \textit{roles}  públicos não devem ter acesso a estas entidades. Note-se que todos os utilizadores com uma conta têm por defeito o \textit{role} \textit{Registered} e todos os utilizadores têm por defeito o \textit{role} \textit{Anonymous}. Estes dois \textit{roles} vão ter a maior \textit{pool} de utilizadores, por esse motivo entidades sensíveis não devem ser acedidas por utilizadores que só possuam \textit{roles} atribuídos por defeito.

\begin{table}[t]
	\centering
	\begin{tabularx}{\columnwidth}{|X|X|X|X|X|X|}
		\hline
		\textbf{} & \textbf{Ecrãs } & \textbf{Ecrãs } & \textbf{Falhas} & \textbf{Falsos} & \textbf{\% de falhas}\\ 
		\textbf{} & \textbf{com falhas} & \textbf{validados} & \textbf{reais} & \textbf{positivos} & \textbf{reais}\\
		\hline
		\textbf{Política 1} & 5 & 5 & 0 & 5 & 0\% \\ \hline
		\textbf{Política 2} & 9 & 9 & 9 & 0 & 100\% \\ \hline
	\end{tabularx}
	
	\caption{Validação dos resultados da fábrica de software A}
	\label{tab:digitalvalidation}
\end{table}

Da fábrica A foram usadas duas políticas de segurança fornecidas pelos donos da fábrica. Executámos a análise e obtemos dois relatórios. Do que observámos, a fábrica A parece estar bem protegida, pois os dois relatórios são pequenos (quando comparados com os da fábrica B). A Tabela~\ref{tab:digitalvalidation} contém os resultados da nossa validação à fábrica A.

As políticas de segurança usadas para a fábrica A foram:

\begin{itemize}
	\item A entidade da \textbf{Política 1} contém informação relacionada com uma aplicação de \textit{feedback} de empregados, e restringe acesso a utilizadores que tenham apenas os \textit{roles} por defeito do sistema.
	\item A entidade da \textbf{Política 2} contém informação relacionada com o sistema de gestão financeira duma empresa. Esta entidade só deve ser acedida pelos \textit{roles} de administrador, \textit{back-office} ou de gestor financeiro.
\end{itemize}

O relatório obtido com a política 1 mostrou 5 ecrãs com falhas de segurança. Todas as 5 falhas reportadas eram falsos positivos, causados por uma limitação na nossa análise. A questão é que a equipa de desenvolvimento da fábrica A utilizam outros mecanismos de segurança para além das primitivas \textit{CheckRole}. Nestes casos o resultado duma query estava a ser filtrado pelo identificador do utilizador. 

O relatório obtido com a política 2 mostrou 9 ecrãs com falhas de segurança. Os 9 ecrãs foram validados e continham falhas de segurança. Estes ecrãs não tinham os \textit{roles} corretos nas suas permissões e não possuíam nenhum mecanismo de segurança nas ações que invocavam.

Da fábrica B foram usadas 5 políticas de segurança, cada uma relacionada com uma entidade na fábrica:

\begin{itemize}
	\item A entidade da \textbf{Política 1} contém informação pessoal sobre os clientes da empresa, e proíbe o seu acesso a utilizadores que só tenham os \textit{roles} por defeito do sistema;
	\item A entidade da \textbf{Política 2} contém informação sobre o sistema de biométrica da empresa,  e proíbe o seu acesso a utilizadores que só tenham os \textit{roles} por defeito do sistema;
	\item A entidade da \textbf{Política 3} contém informação específica sobre empregados da empresa, e proíbe o seu acesso a utilizadores que só tenham os \textit{roles} por defeito do sistema;
	\item A entidade da \textbf{Política 4} contém informação sobre o stock dum produto utilizado na empresa, e proíbe o seu acesso a utilizadores que só tenham os \textit{roles} por defeito do sistema;
	\item A entidade da \textbf{Política 5} contém informação específica sobre alguns clientes da empresa, e proíbe o seu acesso a utilizadores que só tenham os \textit{roles} por defeito do sistema.
\end{itemize}

A Tabela~\ref{tab:atcvalidation} contém os resultados da nossa validação da fábrica B.
Alguns dos relatórios não foram completamente validados, pois tinham muitas entradas ou a complexidade de algumas ações era demasiada alta para conseguir validar sem ter contexto da fábrica de software. 

\begin{table}[!t]
	\footnotesize{
	\centering

	\begin{tabularx}{\textwidth}{|X|X|X|X|X|X|X|}
		\hline
		\textbf{} & \textbf{\textls[-30]{Ecrãs com}} & \textbf{Ecrãs} & \textbf{\%  ecrãs } & \textbf{Falhas} & \textbf{Falsos} & \textbf{\%  falhas} \\
			\textbf{} & \textbf{falhas} & \textbf{validados} & \textbf{validados} & \textbf{reais} & \textbf{positivos} & \textbf{reais} \\ \hline
		\textbf{Política 1} & 280 & 20 & $\approx$ 7\% & 18 & 2 & 90\% \\ \hline
		\textbf{Política 2} & 1  & 1   & 100\% & 1 & 0 & 100\% \\ \hline
		\textbf{Política 3} & 9  & 9   & 100\% & 9 & 0 & 100\% \\ \hline
		\textbf{Política 4} & 14 & 7   & 50\% & 7 & 0 & 100\% \\ \hline
		\textbf{Política 5} & 42 & 8   & $\approx$ 19\% & 8 & 0 & 100\% \\ \hline
	\end{tabularx}}

	\caption{Validação dos resultados da fábrica de software B}
	\label{tab:atcvalidation}
\end{table}

Dos 45 ecrãs com falhas reportadas que validámos, mais de 95\% tinham falhas que necessitaram de ser corrigidas. Estes resultados são muito positivos e mostram que a nossa análise é capaz de encontrar  falhas relevantes em fábricas de software.

A maioria das falhas encontradas foram casos triviais em que um ecrã não tinha as permissões corretas e as ações que chamava não tinham mecanismos de segurança. Estas falhas são fáceis de corrigir, assim que detetadas, mas tendo em conta a dimensão da  fábrica B a sua deteção manual não é exequível. 
Assim sendo, a nossa análise é essencial para  as equipas de desenvolvimento assegurarem a segurança nas suas fábricas de software.

Para exemplificar a gravidade de algumas das falhas encontradas, a falha que encontrámos usando a política 2 comprometia a segurança do sistema de biométrica da fábrica. 
Um dos ecrãs que permitia utilizadores criarem ou alterarem as impressões digitais no sistema não estava restrito ao conjunto certo de \textit{roles} e não existia nenhum mecanismo de segurança nas suas ações.
Isto permitia que um atacante inserisse a sua própria impressão digital no sistema, ou até que removesse ou alterasse a de outro utilizador, fazendo-se passar por ele. 
Validámos esta falha de segurança com os donos da fábrica, que confirmaram a sua gravidade e o desconhecimento da sua existência.

Os 2 falsos positivos detetados devem-se a limitações da nossa análise, referidas na Secção~\ref{sec:solution}. Nestas entradas do relatório estão a ser usados outros mecanismos de segurança que ainda não são contemplados pela nossa análise.

A análise possui algumas limitações.
Nomeadamente, outros mecanismos de segurança para além de \textit{check roles} com condicionais podem ser usados, e a análise ainda não os deteta, resultando em falsos positivos. 
Outra limitação consiste em situações onde a condição dentro dum bloco condicional contém um \textit{check role} em disjunção com outra expressão. 
Condições com disjunções não são triviais e é difícil inferir os  \textit{roles} do utilizador em cada ramo do bloco condicional. 
Nesta situação, a nossa análise segue uma abordagem conservadora, que pode resultar em falsos positivos. Apesar disto, falsos positivos são uma característica comum em análises estáticas, e a existência deles não tira valor às falhas encontradas com sucesso.

\subsection{Performance e Escalabilidade}

Para avaliar a performance da nossa análise, gerámos 20 relatórios utilizando várias políticas de segurança na fábrica B. Para avaliar a performance da nossa análise combinamos políticas definidas especificamente para a avaliação com políticas fornecidas pelos donos da fábrica.

\begin{table}[!t]
    \centering
    \begin{tabular}{|c|c|c|c|c|c|c|}
        \hline
        \textbf{Fábrica} & \textbf{\begin{tabular}[c]{@{}c@{}}Application\\ Objects\end{tabular}} & \textbf{\begin{tabular}[c]{@{}c@{}}Tamanho \\ (Módulos)\end{tabular}} & \textbf{\begin{tabular}[c]{@{}c@{}}Tamanho\\ (MB)\end{tabular}} & \textbf{\begin{tabular}[c]{@{}c@{}}Gerar\\ grafo (s)\end{tabular}} & \textbf{\begin{tabular}[c]{@{}c@{}}Nós\\ grafo\end{tabular}} & \textbf{\begin{tabular}[c]{@{}c@{}}Arcos\\ grafo\end{tabular}} \\ \hline
        \textbf{Mock-up} & 28 & 2 & 2 & \textless{}1 & 83 & 220 \\ \hline
        \textbf{A} & 772 & 34 & 24 & 101 & 3581 & 10352 \\ \hline
        \textbf{B} & 9476 & 691 & 517 & 278 & 48455 & 250122 \\ \hline
    \end{tabular}
    \caption{Resultados de carregar e gerar 3 diferentes fábricas de software}
    \label{tab:gen_graph}
    \vspace{-10pt}
\end{table}

\begin{table}[!t]
	\centering
	\begin{tabular}{|c|c|c|c|}
		\hline
		\textbf{Fábrica} & \textbf{Tamanho (Módulos)} & \textbf{Tamanho (MB)} & \textbf{Espaço em memória (MB)} \\ \hline
		Mock-up & 2 & 2 & $\approx$90\\ \hline
		A & 34 & 25 & $\approx$150\\ \hline
		B & 691 & 517 & $\approx$1024\\ \hline
	\end{tabular}
	\caption{Espaço em memória de 3 diferentes fábricas de software}
	\label{tab:gen_graph_memory}
	\vspace{-10pt}
\end{table}

O tempo de execução médio foi de cerca de 5 minutos, com os piores casos por volta dos 14 minutos. O objetivo desta técnica é gerar um relatório de segurança depois de executar a análise, o qual será posteriormente inspecionado. Como tal, tempos de execução mais longos não são problemáticos

Em termos de escalabilidade, medimos o tempo de carregar e gerar o modelo sintetizado das 3 fábricas, o tamanho resultante dos grafos, e o espaço ocupado em memória pelos modelos sintetizados de cada fábrica. As Tabelas~\ref{tab:gen_graph} e \ref{tab:gen_graph_memory} contém os resultados destas observações. Na segunda coluna da Tabela~\ref{tab:gen_graph}, o termo \textit{Application Objects} é uma métrica usada em OutSystems, que conta o número de páginas e tabelas existentes numa fábrica\footnote{\url{https://success.outsystems.com/Support/Enterprise_Customers/Licensing/Overview/Application_Object_count}}.

Os resultados da nossa análise foram representativos e atingiram os objetivos a que nos propusemos.
Escalabilidade não foi um problema, mesmo ao analisar um fábrica com as dimensões da B. Em ambas as fábricas foram encontradas falhas graves, com um número reduzido de falsos positivos.
Ainda que graves, a correção da maioria das falhas é trivial, sendo o maior desafio a sua deteção manual.

A análise possuí valor, pois vai ajudar as equipas de desenvolvimento a assegurarem a segurança nas suas fábricas de software, reduzindo o tempo despendido no processo de deteção manual de falhas de segurança. Estas equipas de desenvolvimento trabalham sobre plataformas \textit{low-code}, onde podem faltar ferramentas para fazerem testes extensivos. Além disso, mesmo com bons testes é possível não apanhar todas as falhas, logo uma análise estática como a que apresentamos é útil. e Os clientes que forneceram as fábricas  mostraram interesse na análise e esperam poder integrar a análise no processo de desenvolvimento de aplicações.

\section{Trabalho Relacionado}
\label{sec:related}

O modelo de Role-Based Access Control (RBAC)~\cite{sandhu1998role} proposto por Sandhu simplifica a atribuição de permissões a utilizadores dum sistema. Em vez de se atribuírem permissões a utilizadores individuais, criam-se \textit{roles} e atribuem-se \textit{roles} aos utilizadores, e as permissões são dadas a \textit{roles} em vez de utilizadores. 

A nossa análise possuí algumas semelhanças a técnicas de \emph{taint analysis}~\cite{Arzt:2014:FPC:2594291.2594299, newsome2005dynamic, Tripp:2013:AAS:2450312.2450333, Tripp:2009:TET:1543135.1542486}, técnicas que detetam caminhos em que informação não confiável consiga chegar a locais sensíveis onde não devia chegar, sem serem desclassificados. 
É possível fazer uma analogia destes caminhos com aqueles que nossa análise deteta. A diferença é que
\emph{taint analysis} são detetados dados não confiáveis, enquanto a nossa análise deteta \textit{roles} que acedem a indevidamente recursos protegidos.

Mais semelhantes à nossa análise são as técnicas de Model Checking~\cite{clarke2018model, clarke2003counterexample, biere2003bounded, burch1992symbolic, baier2008principles}, que exploram todos os estados possíveis dum sistema de modo a verificar se ele cumpre com um dado conjunto de propriedades. A nossa análise verifica se existe algum estado que não cumpra a política de segurança definida. Da mesma forma que a nossa análise começa por usar um call graph para detetar possíveis falhas, a técnica CEGAR~\cite{clarke2003counterexample}, também começa por utilizar um modelo mais abstrato do sistema, e só quando deteta erros no modelo abstrato é que o refina com mais informação para verificar o erro detetado.

\section{Conclusões}
\label{sec:conc}

Neste artigo apresentámos uma técnica análise de segurança estática que deteta falhas de segurança ao nível de \textit{roles} em fábricas de software.

A técnica foi implementada para fábricas de software OutSystems e detetou com sucesso falhas de segurança em fábricas de software de clientes, algumas delas graves, que dificilmente seriam detetadas sem uma ferramenta de análise. Estas falhas já estavam presentes nas fábricas há bastante tempo e não tinham sido encontradas manualmente. 

Como trabalho futuro, é particularmente importante endereçar outros mecanismos de segurança para além de \emph{check roles} com condicionais e desta forma reduzir o números de falsos positivos.
Planeamos melhorar e aperfeiçoar a ferramenta de modo a que possa ser integrada no ambiente de desenvolvimento OutSystems.
%
%
%
\paragraph{\textbf{Acknowledgements:}}
This work was partially supported by NOVA LINCS (UID/CEC/ 04516/2013) and FCT project PTDC/CCI-INF/32081/2017.

\bibliographystyle{splncs04}
\bibliography{mybibliography}

\end{document}